\documentclass{article}
\usepackage{graphics}

\makeatletter
\@addtoreset{equation}{section}

\makeatother

\begin{document}

\begin{flushright}
May 2008

SNUTP08-002
\end{flushright}

\begin{center}

\vspace{5cm}

{\LARGE 
\begin{center}
Quantum Effects on Tachyon Dynamics
\end{center}
}

\vspace{2cm}

Takao Suyama \footnote{e-mail address : suyama@phya.snu.ac.kr}

\vspace{1cm}

{\it Center for Theoretical Physics, 

Seoul National University, 

Seoul 151-747 Korea}

\vspace{3cm}

{\bf Abstract} 

\end{center}

We investigate the relevance of quantum effects in the study of bulk tachyon dynamics. 
We analyze the behavior of a system which consists of a tachyon, the graviton and the dilaton 
through the use of the Wheeler-DeWitt equation 
in the mini-superspace approximation, 
The singular behavior of the classical solution for the system is modified 
so that the quantum behavior seems to be regular.

\newpage

\vspace{1cm}

\section{Introduction}

\vspace{5mm}

It remains a challenging problem to understand whether a closed string theory with a tachyon is well-defined. 
The case in which the tachyon is localized in the target space is better understood. 
It turns out that the 
tachyon will condense and changes a local geometry of the target space. 
The change may propagate to outer region of the target space, leading to deformed asymptotic structure. 
For a review, see e.g. \cite{review}. 

On the other hand, our understanding of the case with bulk tachyon is very limited. 
The difficulty with this issue can be seen from the analysis of classical solutions. 
In \cite{YZ1}, classical solutions were investigated in 
a low energy effective theory of string theory with a tachyon in the bulk, 
under the assumption of the homogeneity and the isotropy. 
It is observed that all solutions will evolve toward a singular configuration. 
From this observation, one may consider that 
a perturbation around such a solution does not make sense, raising a question whether the full string 
theory with bulk tachyon can make sense. 
The singular behavior of the classical solution was also discussed in \cite{YZ2} from 
the viewpoint of string field theory. 

The analysis in \cite{YZ1} 
is for a string theory in which a tachyon shows up when the theory is expanded around a trivial flat space. 
This implies that the tachyon potential $V(T)$ has a local maximum at, say, $T=0$ at which $V(0)=0$. 
It is possible to consider a supercritical string theory with a tachyon, in which $V(0)>0$, and investigate classical solutions 
similarly 
\cite{FHL}\cite{Suyama}. 
It turns out that the behavior of the classical solution in this case is milder than the critical cases. 
In particular, assuming that $V(T)$ has a minimum at $T=T_0$ at which $V(T_0)>0$, the tachyon rolls down and settles at the 
minimum. 
In the far future, the dilaton approaches a linear function in time, and decreases. 
Therefore, this classical solution could approximate the tachyon dynamics of the full string theory, at least at late stage 
of the time evolution. 
Unfortunately, the classical solution is singular in the past. 
A resolution of this past singularity was proposed in \cite{DDual}\cite{Silverstein1}. 
There is another approach to study condensation of a bulk tachyon when the tachyon profile 
depends only on a light-like coordinate \cite{HS}\cite{HK}\cite{HK2}\cite{Frey}\cite{HS'}\cite{HS4}\cite{HS3}\cite{HS2}. 
In this case, the worldsheet theory is tractable, and one can study the entire process of this class of evolutions. 
There are other researches on closed string tachyon condensation \cite{HS5}\cite{Silverstein2}\cite{HS6}\cite{HL}. 

In this paper, we would like to examine effects of quantum corrections to the classical tachyon evolution. 
We investigate this problem by using the mini-superspace approximation. 
The behavior of the wave function is examined by analyzing 
the Wheeler-DeWitt equation. 
We find an indication that the singular behavior of 
the classical solution is regularized by a quantum effect. 
This is indicated by the decay of the probability density in a region of field space corresponding to the classical singularity. 
The final state of tachyon condensation is, however, not clear from our investigation since the tachyon does not always settle 
at a minimum of $V(T)$.  

Before presenting our investigation, it should be mention that, although we hope to gain some insight on bulk tachyons 
from such an analysis, 
we do not know 
whether this simplification corresponds to a self-consistent approximation in any sense. 
What will be done in this paper is firstly to truncate a string theory with a tachyon to a system which consists 
of the graviton, the dilaton and 
the tachyon, and secondly to reduce the field space by hand. 
The first truncation might be justified for some special cases, for example, when $V(T)$ is slowly-varying in $T$ so that 
any massive string states would not be excited during the evolution. 
However, it is not at all clear how to systematically improve our result to understand the dynamics in string theory. 

Strictly speaking, what we will do is to quantize a system whose classical equations of motion are the same 
as those of the graviton-dilaton-tachyon system reduced according to the ansatz of the homogeneity and the isotropy. 
The integration measure which provides the inner product of wave functions 
should be derivable in principle from the path-integral measure of the full theory. 
However, we simply assume a specific choice so that the Hamiltonian of the reduced system is hermitian with respect to 
the measure. 

It should be noted on the choice of the effective action. 
The choice of the action in this paper is based on the result of \cite{Suyama2} in which a string theory compactified on a 
compact manifold was considered, and a tachyon was supposed to appear from a state which might be localized in the compact 
manifold but which propagated all non-compact directions. 
The word ``bulk tachyon'' in this paper indicates this kind of tachyons. 
If the relation between the equations of motion and the RG flow of the corresponding worldsheet theory is taken into account, 
the form of the effective action is restricted. 
It seems that the same arguments does not apply to the ``universal'' bulk tachyon which corresponds to the unit operator of the 
worldsheet theory. 
There is a proposal of the effective action for the universal tachyon in \cite{Swanson}. 

This paper is organized as follows. 
The behavior of homogeneous and isotropic classical solutions is discussed in section \ref{classical}. 
Generically, the classical solution evolves toward a singular configuration. 
In section \ref{mini}, the Wheeler-DeWitt equation is derived, and its solution is analyzed. 
It is shown that the singular behavior of the classical solution is somehow regularized by a quantum effect. 
It is also shown that the presence of the tachyon tends to regularize the system. 
A possible generalization of our analysis to the situation where the spatial manifold is not flat is briefly 
mentioned in section \ref{curvature}. 
Section \ref{discuss} is devoted to discussion. 
Appendix \ref{E} summarizes some adiabatic properties of a quantum mechanical system.

\vspace{1cm}

\section{Classical solutions} \label{classical}

\vspace{5mm}

In this section, we investigate the behavior of classical solutions of a system specified by the action 
\begin{equation}
S = \frac1{2\kappa^2}\int d^Dx\sqrt{-g}\ e^{-2\Phi}\left[ R+4(\nabla\Phi)^2-(\nabla T)^2-2V(T) \right]. 
   \label{action}
\end{equation}
A class of solutions with the flat metric was investigated in \cite{YZ1}. 
Throughout this paper, the tachyon potential $V(T)$ is assumed to have a global minimum at $T=T_0$. 
Both signs of $V(T_0)$ are allowed. 

\vspace{5mm}

The choice of the action (\ref{action}) we made above is based on the following analysis. 
The details are found in \cite{Suyama2}. 
The most general action of the graviton-dilaton-tachyon system, ignoring all terms with more than two derivatives, is 
\begin{eqnarray}
S 
&=& \frac1{2\kappa^2}\int d^Dx\sqrt{-g}\ e^{-2\Phi}f_1(T)\Bigl[ R+4f_2(T)(\nabla\Phi)^2 
   +f_3(T)\nabla\Phi\cdot\nabla T-f_4(T)(\nabla T)^2-2V(T) \Bigr]. \nonumber \\
   \label{stringframe} 
\end{eqnarray}
We assume that the equations of motion of (\ref{stringframe}) would be obtained by requiring that the beta-functional for the 
conformal invariance of the underlying worldsheet theory vanishes. 
The string theory we consider is assumed to be compactified on a compact manifold $M$, 
and the tachyon $T$ corresponds to a relevant operator of a CFT describing $M$. 
The action of the worldsheet theory is 
\begin{eqnarray}
S_{ws} &=& \frac1{4\pi\alpha'}\int d^2\sigma\sqrt{h}\Bigl[ h^{ab}\partial_a X^\mu\partial_b X^\nu 
 g_{\mu\nu}(X)+\alpha'R^{(2)}\Phi(X) \Bigr] \nonumber \\
& & +S_M(Y)+\int d^2\sigma\sqrt{h}\ T(X)V(Y). 
   \label{worldsheet}
\end{eqnarray}
The action $S_M(Y)$ is that of a CFT 
describing $M$, for which the dynamical variables are collectively denoted by $Y$. 
The operator $V(Y)$ is a vertex operator of the CFT corresponding to the tachyon $T$. 
Suppose that $T(X)$ is a constant. 
Then the non-compact part and the $M$ part of the worldsheet theory (\ref{worldsheet}) 
decouple from each other. 
In this case, the beta-functional $\beta_{\mu\nu}$ 
for $g_{\mu\nu}$, for example, does not depend on $T$ at all. 
This implies that the beta-functional $\beta_{\mu\nu}$ for a generic $T(X)$ should be 
\begin{equation}
\beta_{\mu\nu} = R_{\mu\nu}+2\nabla_\mu\nabla_\nu\Phi +\mbox{ derivatives of }T, 
\end{equation}
at the leading order in $\alpha'$. 
The condition $\beta_{\mu\nu}=0$ and $\beta_\Phi=0$ are consistent with the equations of motion of (\ref{stringframe}) iff 
$f_1(T)=f_2(T)=1$. 
The functions $f_3(T),f_4(T)$ can be eliminated by a field redefinition, and therefore, we obtain (\ref{action}) as the most 
general action suitable for the case we would like to consider. 

Note that in \cite{Swanson} the effective action was determined by requiring that the action allows a simple solution which 
provides an exact CFT. 
In our case, since we would like to consider a tachyon which corresponds to a non-trivial relevant operator, we should not expect 
the existence of such a simple solution. 
Therefore, the effective action (\ref{action}) may differ from the one used in \cite{Swanson}. 

\vspace{5mm}

The equations of motion derived from (\ref{action}) are 
\begin{eqnarray}
R_{\mu\nu}+2\nabla_\mu\nabla_\nu-\nabla_\mu T\nabla_\nu T &=& 0, \nonumber \\
\nabla^2\Phi-2(\nabla\Phi)^2-V(T) &=& 0, \\
\nabla^2T-2\nabla\Phi\cdot\nabla T-V'(T) &=& 0. \nonumber 
\end{eqnarray}
We make the following ansatz 
\begin{eqnarray}
ds^2 &=& -dt^2+e^{-2B(t)}\delta_{ij}dx^idx^j, \nonumber \\
\Phi &=& \Phi(t), 
   \label{ansatz} \\
T &=& T(t). \nonumber 
\end{eqnarray}
The equations of motion then reduce to 
\begin{eqnarray}
\ddot{B}-((D-1)\dot{B}+2\dot{\Phi})\dot{B} &=& 0, \nonumber \\
\ddot{\Phi}-((D-1)\dot{B}+2\dot{\Phi})\dot{\Phi}+V(T) &=& 0, 
    \label{eom} \\
\ddot{T}-((D-1)\dot{B}+2\dot{\Phi})\dot{T}+V'(T) &=& 0, \nonumber 
\end{eqnarray}
with a constraint 
\begin{equation}
(D-1)(D-2)\dot{B}^2+4(D-1)\dot{B}\dot{\Phi}+4\dot{\Phi}^2-\dot{T}^2-2V(T) = 0, 
   \label{constraint0}
\end{equation}
which just restricts the initial condition of this system. 
There is a simple integral of motion 
\begin{equation}
p_c = 2(D-1)e^{-(D-1)B-2\Phi}\dot{B}.
\end{equation}
The coefficient is chosen for later convenience. 

It is convenient to introduce 
\begin{equation}
K = (D-1)\dot{B}+2\dot{\Phi}. 
\end{equation}
In terms of these variables, the equations of motion can be written as  
\begin{eqnarray}
\ddot{B}-K\dot{B} &=& 0, \\
\dot{K}-K^2+2V(T) &=& 0, \\
\ddot{T}-K\dot{T}+V'(T) &=& 0, 
\end{eqnarray}
with 
\begin{equation}
K^2-(D-1)\dot{B}^2-\dot{T}^2-2V(T) = 0. 
   \label{constraint}
\end{equation}
It is easy to see that $\dot{K}$ is always non-negative due to the constraint (\ref{constraint}) since 
\begin{eqnarray}
\dot{K} &=& K^2-2V(T) \nonumber \\
&=& (D-1)\dot{B}^2+\dot{T}^2 \ \ge\ 0. 
\end{eqnarray}
The equality holds only when $\dot{B}=\dot{T}=0$. 

The behavior of classical solutions strongly depends on the sign of $V(T_0)$. 
Let us investigate them for each case. 

\vspace{5mm}

\subsection{$V(T_0)<0$}

\vspace{5mm}

This case would be interesting since the analysis of this situation would have some implications to the 
ground state of a string theory with a tachyon which is perturbatively defined in the flat space-time. 

The simplest solution is obtained by setting $T=T_0$. 
There is no time-independent solution which is suitable for a vacuum. 
Note that the space-like linear dilaton solution, which is a well-known solution to (\ref{action}), is not compatible with our 
ansatz (\ref{ansatz}), and therefore, they are not discussed in this paper. 
The general solution in this case is 
\begin{eqnarray}
B &=& \frac{1}{2\sqrt{D-1}}\log\frac{1+\sin a(t-t_0)}{1-\sin a(t-t_0)}+B_0, 
   \label{soln1} \\
\Phi &=& -\frac14\left( 1+\sqrt{D-1} \right)\log(1+\sin a(t-t_0)) \nonumber \\ 
& &-\frac14\left( 1-\sqrt{D-1} \right)\log(1-\sin a(t-t_0))+\Phi_0,  
   \label{soln2}
\end{eqnarray}
where $t_0,\Phi_0$ and $B_0$ are integration constants, and $a=\sqrt{-2V(T_0)}$. 
The constant of motion $p_c$ can be written 
in terms of these integration constants as  
\begin{equation}
p_c^2 = 8(D-1)|V(T_0)|e^{-2(D-1)B_0-4\Phi_0}. 
   \label{p_c}
\end{equation}
The solution always evolves from a strongly coupled background at a finite past to a big crunch at a finite future. 
There is also the time-reversal solution which evolves from a big bang to a strongly coupled background. 
In both cases, the classical solutions have singular behaviors. 
In fact, one can show that the energy-momentum tensor (defined for the Einstein frame metric) diverges when the background 
becomes singular, and therefore, they are physical singularities. 
Both $B$ and $\Phi$ diverge in a finite time, while keeping either $(D-1)B+2\Phi+\sqrt{D-1}B$ or $(D-1)B+2\Phi-\sqrt{D-1}B$ 
finite since 
\begin{equation}
(D-1)B+2\Phi\pm\sqrt{D-1}B = -\log(1\mp\sin a(t-t_0)) + \mbox{const.}
   \label{behavior0}
\end{equation}

The singular behavior of the classical solution persists when the dynamics of the tachyon $T$ is included. 
Assuming that there is no extremum of $V(T)$ above zero, $\dot{K}$ cannot approach toward 0 in the limit $t\to+\infty$, 
implying that $K\to+\infty$. 
Note that this is a generic behavior even when there are extrema above zero, except that the initial condition is fine tuned. 
The divergence of $K$ implies that $\dot{B}\to+\infty$ or $\dot{\Phi}\to+\infty$ will be realized. 
The former is a solution with the big crunch, while the latter solution evolves into a strongly coupled one. 
In any case, the classical description would break down there, and 
some quantum effects due to the high curvature or the string loops would become important in this limit. 

\vspace{5mm}

\subsection{$V(T_0)>0$}   \label{V>0}

\vspace{5mm}

Tachyon condensation in this case is discussed in \cite{FHL}\cite{Suyama}. 
In this case, the time evolution of this system is more regular than in the previous case. 

For the case in which $T=T_0$, there are two solutions. 
The first solution is the well-known time-like linear dilaton background. 
The second one is 
\begin{eqnarray}
B &=& \frac1{\sqrt{D-1}}\log \tanh\frac{\alpha}2(t-t_0) + B_0, \\
\Phi &=& -\log \sinh \alpha(t-t_0)-\sqrt{D-1}\log \tanh\frac{\alpha}2(t-t_0) + \Phi_0, 
\end{eqnarray}
where $\alpha=\sqrt{2V(T_0)}$. 
Note that this solution is valid for $t>t_0$. 
Near $t=t_0$, the string coupling diverges, while keeping 
\begin{equation}
(D-1)B+2\Phi+\sqrt{D-1}B = -\log(1+\cosh \alpha(t-t_0)) + \mbox{const.}
   \label{behavior}
\end{equation}
finite. 
At later time, the string coupling becomes smaller, and $B$ approaches a constant value $B_0$. 
Therefore, this solution is regular in the future. 
In fact, this solution approaches the linear dilaton solution. 
The expression for $p_c$ in this case is the same as (\ref{p_c}). 

The qualitative behavior of the solution including the dynamical tachyon can be easily understood as follows. 
The constraint (\ref{constraint}) implies $K^2\ge 2V(T_0)$. 
It is enough to consider only the case $K\le-\sqrt{2V(T_0)}$, since the other case leads to the time reversal solution. 
The equation of motion of the tachyon leads 
\begin{equation}
\frac d{dt}\left[ \frac12\dot{T}^2+V(T) \right] = K\dot{T}^2. 
\end{equation}
This implies that the ``energy'' of the tachyon decreases until $\dot{T}=0$ is realized. 
This can happen at a (local) minimum of $V(T)$. 
Therefore, whatever the initial condition is, the tachyon settles down at a minimum of the potential, and the background 
approaches a linear dilaton, in the future.

\vspace{1cm}

\section{Mini-superspace analysis}  \label{mini}

\vspace{5mm}

We have shown that any classical solution of a homogeneous and isotropic rolling tachyon evolves from or 
toward a singular configuration (possibly in a 
finite time).  
A natural question is whether this classical behavior would be modified by some quantum effects. 
Of course, it is almost impossible, at least 
up to now, to take into account the quantum corrections in quantum gravity and string theory, 
except for some limited cases with large symmetry groups. 
What we are going to investigate below is to consider the quantum effects in simplified models. 
This is known as the mini-superspace approximation. 

The equations of motion (\ref{eom}) as well as the constraint (\ref{constraint0}) can be derived from the action 
\begin{eqnarray}
S_R &=& \int dt\left[ e^{A-(D-1)B-2\Phi}\left\{ -(D-1)(D-2)\dot{B}^2-4(D-1)\dot{\Phi}\dot{B}-4\dot{\Phi}^2+\dot{T}^2 \right\}
 \right.   \nonumber \\
& & \left. -2\ e^{-A-(D-1)B-2\Phi}V(T) \right]. 
\end{eqnarray}
This action can be obtained, up to an overall constant, 
from the action (\ref{action}) by substituting the ansatz (\ref{ansatz}) in which the metric is 
replaced with 
\begin{equation}
ds^2 = -e^{-2A(t)}dt^2+e^{-2B(t)}\delta_{ij}dx^idx^j. 
\end{equation}
Note that the variation with respect to $A$ provides the constraint (\ref{constraint0}). 
We expect that the quantum system specified by the reduced action $S_R$ could be regarded as an approximation to the full 
quantum gravity or string theory. 

Since $S_R$ describes a one-dimensional system, it is rather straightforward to quantize this system. 
This system is invariant under the reparametrization of the time coordinate, and therefore, 
it is possible to choose the time coordinate so that $A(t)=0$. 
The remnant of this reparametrization invariance is the so-called Hamiltonian constraint imposed 
on the wave function. 
Quantum mechanically, it is given as 
\begin{equation}
\left[ \frac1{16}\partial_\phi^2-\frac1{4(D-1)\phi^2}\partial_B^2-\frac1{4\phi^2}\partial_T^2+2\phi^2V(T) 
 \right] \Psi(\phi,B,T) = 0, 
    \label{WDW}
\end{equation}
where 
\begin{equation}
\phi = e^{-\frac12(D-1)B-\Phi}. 
\end{equation}
This equation is known as the Wheeler-DeWitt equation. 
It is interesting to observe that $\phi$ looks like a time coordinate since the sign of the kinetic term is opposite to the 
other two variables. 
This is in harmony with that fact that, in the classical solutions, $(D-1)B+2\Phi$ never decreases which enables us to identify 
it with the time. 

To extract some physical information from the wave function $\Psi(\phi,B,T)$, it is necessary to define 
the integration measure, or the inner product, for the wave function. 
The inner product should be defined such that the Hamiltonian operator appeared in (\ref{WDW}) is hermitian. 
In the following, we simply assume that the measure is the trivial one, so that $|\Psi(\phi,B,T)|^2$ provides us the probability 
for the variables $\phi,B,T$ to have some specified values. 
Note that the probability for $\Phi,B,T$ is then given by $e^{-\frac{D-1}2B-\Phi}|\Psi(e^{-\frac{D-1}2B-\Phi},B,T)|^2$. 

\vspace{5mm}

\subsection{Fixed $T$}

\vspace{5mm}

Before analyzing the solution to the equation (\ref{WDW}), it is instructive to consider a further simplified equation 
corresponding to the situation with $T=T_0$. 
The following analysis is a straightforward generalization of the one for the 
two-dimensional dilaton gravity theory \cite{2dim}. 
Note that in two dimensions the analysis can be regarded as an exact one. 

The choice $T=T_0$ is consistent with the reduction, and therefore, the Wheeler-DeWitt equation 
\begin{equation}
\left[ \frac1{16}\partial_\phi^2-\frac1{4(D-1)\phi^2}\partial_B^2+2V(T_0)\phi^2 \right] \Psi(\phi,B) = 0, 
\end{equation}
can be derived in the similar manner. 

The wave function can be expanded as 
\begin{equation}
\Psi(\phi,B) = \sum_p c_pe^{ipB}\psi_p(\phi), 
\end{equation}
where $\psi_p(\phi)$ satisfies 
\begin{equation}
\left[ \frac1{16}\partial_\phi^2+\frac{p^2}{4(D-1)\phi^2}+2V(T_0)\phi^2 \right] \psi_p(\phi) = 0. 
   \label{eqpsi}
\end{equation}
This is a one-dimensional Schr\"odinger equation with a non-trivial potential 
\begin{equation}
U(\phi) = -\frac{p^2}{4(D-1)\phi^2}-2V(T_0)\phi^2
\end{equation}

In the case $V(T_0)>0$, the potential $U(\phi)$ is unbounded from below in the large $\phi$ region. 
This reflects the fact that in this case the classical solution approaches the linear dilaton background, and therefore 
$\phi$ grows without bound. 
It will be shown below that this unboundedness of $U(\phi)$ does not cause any strange behavior of the wave function. 

The behavior of $U(\phi)$ near $\phi=0$ is also anticipated from the analysis of classical solutions. 
Since $U(\phi)$ is also unbounded in this region, the wave function generically tends to gather around $\phi=0$ which corresponds 
to the singular background. 
However, this will be modified by quantum effects. 
In fact, the solution of (\ref{eqpsi}) is obtained in the WKB approximation as 
\begin{equation}
\phi_p(\phi,B) \sim U(\phi)^{-\frac14}e^{4i\int^\phi dx\sqrt{-U(x)}}. 
\end{equation}
Here the prefactor $U(\phi)^{-\frac14}$, which is a quantum correction to $\psi_p(\phi,B)$, behaves as $\sqrt{\phi}$ near 
$\phi=0$. 
In terms of $B$ and $\Phi$, this suggests that the probability for the large positive $(D-1)B+2\Phi$ region is exponentially 
suppressed. 
Recalling that this region corresponds to the singularity in the classical solution, irrespective of the sign of $V(T_0)$, 
it is expected that the singular behavior of the classical solution may be regularized in some manner by a quantum effect. 

It turns out that this WKB analysis provides the right picture of the quantum dynamics of the system. 
The Schr\"odinger equation (\ref{eqpsi}) can be solved exactly. 
For the case $V(T_0)<0$, the solution is given by the modified Bessel function as 
\begin{equation}
\psi_p(\phi) = \sqrt{\phi}\ K_{\nu(p)}(2a\phi^2), 
\end{equation}
where $\nu(p)$ is defined as 
\begin{equation}
\nu(p)^2 = -\frac1{D-1}\left( p^2-\frac{D-1}{16} \right). 
\end{equation}
The integration constants are chosen such that $\psi_p(\phi)$ vanishes for large $\phi$, which reflects the behavior of the 
classical solution in which $(D-1)B+2\Phi$ is bounded from below. 

We have obtained the following exact solution of the Wheeler-DeWitt equation: 
\begin{equation}
\Psi(\phi,B) = \sqrt{\phi}\int_{-\infty}^{+\infty} dp\ c(p)e^{ipB}K_{\nu(p)}(2a\phi^2). 
\end{equation}
The function $c(p)$ should be determined in some manner for which the space-time described by $\Psi(\phi,B)$ 
satisfies (a quantum version of) the Einstein equation. 
Here, we assume a qualitative behavior of $c(p)$ based on the following reasoning. 
Recall that $p$ is the eigenvalue of the canonical momentum $\pi_B$ of $B$, which is explicitly given as 
\begin{equation}
\pi_B = 2(D-1)\phi^2\dot{B}. 
\end{equation}
Therefore, for the wave function which is a quantum description of the classical solution, the function $c(p)$ should have 
a sharp peak at $p=p_c$. 
We assume $c(p)\sim e^{-m(p-p_c)^2}$ near $p=p_c$. 
It would be relevant to consider the situation where at an instant of time the spatial manifold has a finite size and the string 
coupling is small. 
In this situation, $p_c$ can be taken to be a large value. 

The behavior of $\Psi(\phi,B)$ near $\phi=0$ is analyzed as follows. 
Recall that the modified Bessel function behaves as a linear combination of plane waves 
$e^{\pm i\frac p{\sqrt{D-1}}((D-1)B+2\Phi)}$ for large $p$. 
Then, the wave function behaves as 
\begin{equation}
\Psi(\phi,B) \sim I_+(\phi,B)+I_-(\phi,B), 
\end{equation}
where 
\begin{equation}
I_{\pm}(\phi,B) \sim e^{-\frac14((D-1)B+2\Phi)}e^{-\frac1{4m(D-1)}\left[ (D-1)B+2\Phi\pm\sqrt{D-1}B) \right]^2}. 
\end{equation}
Without the ``quantum'' factor $\sqrt{\phi}$, $\Psi(\phi,B)$ decays rapidly in the small $\phi$ region unless 
$(D-1)B+2\Phi\pm\sqrt{D-1}B$ is kept finite. 
This precisely describes the behavior (\ref{behavior0}) of the classical solutions. 
And, as suggested by the WKB analysis, the factor $\sqrt{\phi}$ forbids $\phi$ to becomes small, implying a regularization of 
the singular behavior. 
Note that, although $\sqrt{\phi}$ should be the result of a quantum effect, its origin in view of the full theory is not clear. 

For the case $V(T_0)>0$, $\psi_p(\phi)$ is given by the Bessel function as 
\begin{equation}
\psi_p(\phi) = A_p \sqrt{\phi}J_{\nu(p)}(2\alpha\phi^2)+B_p \sqrt{\phi}J_{-\nu(p)}(2\alpha\phi^2). 
\end{equation}
For any choice of $A_p$ and $B_p$, $\psi_p(\phi)$ decays at least as fast as $\phi^{-\frac12}$ 
in the large $\phi$ region, although 
the potential $U(\phi)$ is unbounded from below there. 
In this case, the coefficients can be determined by examining the behavior of $\Psi(\phi,B)$ in the small $\phi$ region, 
and comparing it with the classical solution. 
By the same argument as in the case $V(T_0)<0$, $\Psi(\phi,B)$ behaves as 
\begin{equation}
\Psi(\phi,B) \sim A_{p_c}I_+(\phi,B)+B_{p_c}I_-(\phi,B). 
\end{equation}
Recalling the behavior (\ref{behavior}), the proper choice will be $B_p=0$. 
Classical singularity at $\phi=0$ is also regularized in the same way. 
Note that the normalizability of $\Psi(\phi,B)$ in this case might be problematic since it is naively expected that 
$\Psi(\phi,B)$ behaves as $\phi^{-\frac12}$ in the large $\phi$ region, and we assumed that the inner product is defined by 
the trivial measure. 
It may be appropriate that in this case the wave function should be normalized in the similar manner with plane waves. 
In any case, to clarify this issue, it would be necessary to determine a more detailed form of $c(p)$ to find more detailed 
behavior of the wave function.

\vspace{5mm}

\subsection{Dynamical $T$}  \label{dynamical}

\vspace{5mm}

We have seen that the quantum-corrected solution seems to behave better than the classical one. 
To regard this result as meaningful to the full theory, 
it is necessary to check at least whether the behavior of the wave function $\Psi(\phi,B)$ discussed in the previous subsection 
persists when the dynamics of $T$ is included. 
Recall that the Wheeler-DeWitt equation including $T$ is 
\begin{equation}
\left[ \frac1{16}\partial_\phi^2-\frac1{4(D-1)\phi^2}\partial_B^2-\frac1{4\phi^2}\partial_T^2+2\phi^2V(T) \right]
 \Psi(\phi,B,T) = 0. 
\end{equation}
As in the previous subsection, the wave function can be expanded as 
\begin{equation}
\Psi(\phi,B,T) = \sum_p c_pe^{ipB}\psi_p(\phi,T), 
\end{equation}
and $\psi_p(\phi,T)$ satisfies 
\begin{equation}
\left[ \frac1{16}\partial_\phi^2+\frac{p^2}{4(D-1)\phi^2}-\frac1{4\phi^2}\partial_T^2+2\phi^2V(T) \right]
 \psi_p(\phi,T) = 0. 
\end{equation}
As in the previous subsection, $c_p$ is assumed to be a function of $p$ which is sharply peaked at $p=p_c$, and $p_c$ is 
assumed to be large. 

It turns out to be convenient to define 
\begin{equation}
\psi_p(\phi,T) = e^{\frac \tau 2}\varphi_p(\tau,T), 
\end{equation}
where $\phi=e^\tau$. 
Then $\varphi_p(\tau,T)$ satisfies 
\begin{equation}
\left[ \partial_\tau^2-4\partial_T^2+m(p)^2+32e^{4\tau}V(T) \right] \varphi_p(\tau,T) = 0, 
   \label{WDWwithT}
\end{equation}
where 
\begin{equation}
m(p)^2 = \frac{4p^2}{D-1}-\frac14 \ \left(\ =-4\nu(p)^2\ \right). 
\end{equation}

In the region where $\phi$ is small, or $\tau$ is large and negative, the ``potential term'' $32e^{4\tau}V(T)$ will be negligible, 
and the equation (\ref{WDWwithT}) reduces to the two-dimensional Klein-Gordon equation, in which $\tau$ is the time coordinate 
and $T$ is the spatial coordinate. 
$p$ is assumed to take a large value, and therefore, this is the equation for a massive field. 

The ``time'' evolution of $\varphi_p(\tau,T)$ in this region can be easily deduced as follows. 
If the tachyon potential behaves as $V(T)\to+\infty$ in the limit $T\to\pm\infty$, then at a finite $\tau$ the wave 
$\varphi_p(\tau,T)$ would be confined in a finite region in the $T$-direction. 
When the ``potential term'' becomes negligible, the wave can spread in the $T$-direction, and therefore, the magnitude 
of $\varphi_p(r,T)$ would decrease as $\tau$ becomes large and negative. 
Recall that the wave function $\Psi(\phi,B,T)$ contains $\sqrt{\phi}=e^{\frac \tau2}$ in addition to $\varphi_p(\tau,T)$. 
As a result, the wave function decays rapidly in the direction $2\tau=-(D-1)B-2\Phi$, faster than the case for a constant $T$. 
Therefore, the regularization mechanism seems to persist in the system with the dynamical tachyon. 

To see the behavior of the wave function in more detail, let us make the following 
ansatz: 
\begin{equation}
\varphi_p(\tau,T) = \sum_n d_n(\tau)f_n(T;\tau), 
\end{equation}
where $f_n(T;\tau)$ satisfy
\begin{equation}
\left[ -\frac12\partial_T^2+4e^{4\tau}v(T) \right]f_n(T;\tau) = E_n(\tau)f_n(T;\tau). 
   \label{SchT}
\end{equation}
Here $v(T)$ is defined as 
\begin{equation}
v(T) = V(T)-V(T_0)\ \ge\ 0. 
\end{equation}
The functions $f_n(T;\tau)$ are the wave functions of the tachyon in the non-negative potential $v(T)$ for a fixed $\tau$. 
They are assumed to form an orthogonal basis, that is, 
\begin{equation}
\int dT\ f_m(T;\tau)f_n(T;\tau) = N_m(\tau)\delta_{mn}. 
\end{equation}
Then, the Wheeler-DeWitt equation (\ref{WDWwithT}) reduces to the following set of ordinary differential equations 
\begin{equation}
\left[ -\partial_\tau^2-m(p)^2-32V(T_0)e^{4\tau}-8E_n(\tau) \right]d_n(\tau) = 
 2\sum_m\beta_{nm}(\tau)\partial_\tau d_m(\tau)+\sum_m\gamma_{nm}(\tau)d_m(\tau),    \label{d_n}
\end{equation}
where 
\begin{eqnarray}
\beta_{nm}(\tau) &=& \frac1{N_n(\tau)}\int dT\ f_n(T;\tau)\partial_\tau f_m(T;\tau), \\
\gamma_{nm}(\tau) &=& \frac1{N_n(\tau)}\int dT\ f_n(T;\tau)\partial_\tau^2f_m(T;\tau). 
\end{eqnarray}

It can be found from (\ref{d_n}) that the presence of the dynamical tachyon induces another interesting effects. 
The coefficient functions $d_n(\tau)$ correspond to $\psi_p(\phi)$ in the previous subsection. 
The effective potential for $d_n(\tau)$ is 
\begin{equation}
U_n(\tau) = -m(p)^2-32V(T_0)e^{4\tau}-8E_n(\tau). 
\end{equation}
For the constant $T$ case, the last term is absent. 
From the definition (\ref{SchT}), it is obvious that $E_n(\tau)$ is a non-negative function. 
Moreover, it is a monotonically increasing function of $\tau$, and grows slower than $e^{4\tau}$ in the large $\tau$ region. 
Therefore, the presence of the tachyon makes the wave function to distribute in a large positive 
$\tau$ region, while the asymptotic behavior of the potential in the limit $\tau\to\pm\infty$ is kept intact. 
Details of the behavior of $E_n(\tau)$ are summarized in Appendix \ref{E}. 
This would also be an indication that the dynamics of the tachyon tends to regularize the system. 

The equations (\ref{d_n}) also show that all quantum states of the tachyon couple among them. 
Therefore, even if the tachyon stays in the ground state at, say, $\tau=+\infty$, any other states may be excited in a 
finite $\tau$ region. 
Note that this behavior is expected from the classical solutions. 
For the case $V(T_0)<0$, we should regard $-\tau$ as the time coordinate, at least in the later stage of the evolution. 
The coupling among the tachyon states corresponds to the amplification of the tachyon motion due to the ``Hubble parameter'' 
$K$. 
For the case $V(T_0)>0$, on the other hand, $\tau$ should be regarded as the time. 
In this case, as shown in subsection \ref{V>0}, 
the oscillation of the tachyon around $T=T_0$ damps due to $K$ (note that $K$ is 
always negative). 
Viewing in the opposite direction, this is an amplification of the oscillation. 

To study the region $\tau\to+\infty$, let us consider the simplest potential  
\begin{equation}
v(T) = \frac12\omega^2e^{4r}T^2, 
\end{equation}
where $T$ has been shifted so that the minimum of $v(T)$ is now at $T=0$. 
In fact, by the rescaling $T\to e^{-2\tau}T$ and taking the limit $\tau\to+\infty$, the Schr\"odinger equation (\ref{SchT}) 
reduces to this case as long as $v''(0)>0$ is satisfied. 
Therefore, the analysis of this simplest case would provide a good approximation in the large positive $\tau$ region. 

In this case, the functions $f_n(T;\tau)$ can be written as 
\begin{equation}
f_n(T;\tau) = H_n(\sqrt{\omega}e^\tau T)e^{-\frac\omega2e^{2\tau}T^2}, 
\end{equation}
where $H_n(x)$ are Hermite polynomials. 
The energy is 
\begin{equation}
E_n(\tau) = \omega\left( n+\frac12 \right)e^{2\tau}. 
\end{equation}

It turns out that the matrices $\beta_{nm}(\tau)$ and $\gamma_{nm}(\tau)$ are independent of $\tau$. 
Explicitly, 
\begin{eqnarray}
\beta_{nm} &=& -\frac14\delta_{n,m+2}-\frac12\delta_{nm}+\frac12m(m-1)\delta_{n,m-2}, \\
\gamma_{nm} &=& \sum_{l}\beta_{nl}\beta_{lm}. 
\end{eqnarray}
Since these are constants, it is now apparent that, in the large positive $\tau$ region, the equations (\ref{d_n}) are 
approximated as 
\begin{equation}
\left[ -\partial_\tau^2-32V(T_0)e^{4\tau} \right]d_n(\tau) \sim 0. 
\end{equation}
The functions $d_n(\tau)$ decay very rapidly for a positive $V(T_0)$, while they oscillate for a negative $V(T_0)$. 
The important point is that $d_n(\tau)$ becomes independent of $n$ up to the overall coefficients. 
As a result, the wave function $\varphi_p(\tau,T)$ behave as 
\begin{equation}
\varphi_p(\tau,T) \sim d(\tau)e^{-\frac\omega2e^{2\tau}T^2}\sum_n d_nH_n(\sqrt{\omega}e^\tau T). 
\end{equation}
For a large positive $\tau$, the wave function is localized around $T=0$ in the $T$-direction, whose width becomes narrower 
as $\tau$ increases. 
For the case $V(T_0)>0$, this precisely describes the dynamics of the tachyon as seen from the classical solutions. 

Note that the behavior of the wave function in this region is almost the same as that for the constant $T$ case discussed 
in the previous subsection, except for the presence of the non-trivial tachyon wave function. 
This is actually an expected result. 
What has been done above is an adiabatic approximation. 
From the Wheeler-DeWitt equation, it is obvious that, if $\tau$ is large and positive, 
the tachyon will oscillate very quickly due to the factor $e^{4\tau}$ in front of $V(T)$. 
Therefore, it is possible to regard the coordinate $\tau$ as a slowly-varying parameter as far as 
the tachyon dynamics is concerned.

\vspace{1cm}

\section{Curved spatial manifolds}  \label{curvature}

\vspace{5mm}

It is possible to generalize our analysis so far by employing the following metric form 
\begin{equation}
ds^2 = -e^{-2A(t)}dt^2+e^{-2B(t)}h_{ij}(x)dx^idx^j, 
\end{equation}
where $h_{ij}(x)$ is an Einstein metric 
\begin{equation}
R_{ij} = \lambda h_{ij}. 
\end{equation}
Our analysis so far corresponds to the case $\lambda=0$. 
The modification appears in the equations of motion of $B$ as 
\begin{equation}
\ddot{B} = K\dot{B}+\lambda e^{2B}, 
\end{equation}
while the equations for $\Phi$ and $T$ are kept intact. 
Now there is no simple integral of motion. 
$B$ evolves toward $+\infty$ when $\lambda$ is positive, and toward $-\infty$ when $\lambda$ is negative. 

The equation of $K$ and the constraint are 
\begin{eqnarray}
\dot{K}-K^2+2V(T)-(D-1)\lambda e^{2B} &=& 0, \\
K^2-2V(T)+(D-1)\lambda e^{2B}-(D-1)\dot{B}^2-\dot{T}^2 &=& 0. 
\end{eqnarray}
These imply that $(D-1)B+2\Phi$ can still be regarded as a time coordinate since it is a monotonic function of time. 

The Wheeler-DeWitt equation is 
\begin{equation}
\left[ -\partial_\tau^2+\frac4{D-1}\partial_B+4\partial_T^2+\frac14+16e^{4\tau}\left( -2V(T)+(D-1)\lambda e^{2B} \right) \right]
 \Psi(\tau,B,T) = 0. 
\end{equation}
A quite different phenomenon would happen when $\lambda$ is positive. 
As in subsection \ref{dynamical}, one can first solve 
\begin{equation}
\left[ -\partial_B^2-16(D-1)\lambda e^{4\tau+2B} \right]g_n(B;\tau) = \varepsilon_n(\tau)g_n(B;\tau), 
\end{equation}
and then solve the resulting equations. 
If $\lambda$ is positive, then $\varepsilon_n(\tau)$ can be negative. 
Note that in this case a suitable boundary condition e.g. the one discussed in \cite{Sbrane} should be specified. 
Such negative energy states would result in two-dimensional wave equations in the $\tau$-$T$ space 
with a ``tachyonic'' mass term, and therefore, 
the wave function $\Psi(\tau,B,T)$ may not decay in the large negative $\tau$ region. 
This seems to imply that the expected classical singularity might not be regularized for the case $\lambda>0$. 
It would be very interesting to investigate the behavior of the wave function for this generalized system.

\vspace{1cm}

\section{Discussion}   \label{discuss}

\vspace{5mm}

We have investigated the mini-superspace analysis of the dynamics of a bulk tachyon coupled to background massless fields. 
It is observed that the singular behavior in the classical solutions is regularized in the ``quantum solutions'' described by 
the solutions of the Wheeler-DeWitt equation. 
It is interesting to notice that the dynamics of the tachyon also contributes to regularizing the system. 
The mechanism of this regularization in terms of the full theory is, however, not yet clear. 
It would be very interesting to investigate this mechanism further. 

Our analysis does not clarify what would be the possible final state of bulk tachyon condensation for the case $V(T_0)<0$. 
The case $V(T_0)>0$ was analyzed in section \ref{mini}, and the result agrees with the one expected from the classical solution. 
In the case $V(T_0)<0$, one would like to consider the situation in which the tachyon is localized at a maximum of $v(T)$ 
at $\tau\to+\infty$, 
but if so, the quadratic approximation for $v(T)$ is not valid. 
A possible guess based on our analysis would be that, since all the tachyonic states couple among them, they are ``thermalized'' 
at a finite $\tau$ which characterizes a possible final state of the time evolution. 
This would be a quite different state from the one expected from an analogy with the known examples of tachyon condensations. 
For example, $\langle T \rangle$ might vanish while the vevs of higher powers of $T$ might not. 

The interpretation of the wave function we obtained needs better understanding. 
The amplitude of the wave function decays in some region, as we have shown. 
This can be regarded as a regularization of the classical singularity. 
However, this also seems to suggest that there is a bound on the time coordinate. 
One possible interpretation of this is that $(D-1)B+2\Phi$ would approach a finite constant value in the limit $t\to+\infty$ or 
$t\to t_0$. 
This would be possible since quantum corrections should modify the equations of motion, and therefore, the $t$-dependence of the 
fields should also be modified. 
Another interpretation would be that this may suggest a bounce, and the system then starts evolving oppositely in the 
$(D-1)B+2\Phi$ direction. 
This process, if possible, might look similar to the phenomenon discussed in \cite{CHT} in the context of AdS/CFT correspondence. 

In any case, our analysis suggests that quantum corrections would be very important to obtain a reasonable picture of bulk 
tachyon condensation. 
Despite the difficulty in investigating this issue in a reliable manner, it is worth studying this issue further. 

\vspace{2cm}

{\bf \Large Acknowledgments}

\vspace{5mm}

I would like to thank Soo-Jong Rey and Ian Swanson for valuable comments. 
This work is supported in part by the Korea Research Foundation Leading Scientist Grant (R02-2004-000-10150-0), 
Star Faculty Grant (KRF-2005-084-C00003) and the Korea Research Foundation Grant, No. KRF-2007-314-C00056.

\newpage

\appendix

\vspace{1cm}

{\bf \LARGE Appendix}

\vspace{5mm}

\section{$E_n(\tau)$}  \label{E}

\vspace{5mm}

Consider the Schr\"odinger equation 
\begin{equation}
\left[ -\frac{d^2}{dx^2}+e^{\lambda}V(x) \right]\psi_n(x;\lambda) = E_n(\lambda)\psi_n(x;\lambda). 
\end{equation}
The wave functions are assumed to form an orthonormal basis of the corresponding Hilbert space. 
In the following, it is assume that $V(x)$ is non-negative and vanishes at its minimum. 

The energy $E_n(\lambda)$ can be written as 
\begin{equation}
E_n(\tau) = \int dx\ \psi_n^*(x;\lambda)\left[ -\frac{d^2}{dx^2}+e^{\lambda}V(x) \right]\psi_n(x;\lambda). 
\end{equation}
Obviously, $E_n(\lambda)$ is positive. 
Its derivative is 
\begin{equation}
\frac{dE_n}{d\lambda} = e^{\lambda}\int dx\ |\psi_n(x;\lambda)|^2V(x)\ \ge\ 0. 
   \label{E'}
\end{equation}
The integral in the RHS is estimated as follows. 
As $\lambda$ grows, the wave functions are more and more localized around the global minima of $V(x)$, and the width 
of the peak of $|\psi_n(x;\lambda)|^2$ around each 
minimum becomes smaller and smaller. 
Therefore, the integral would vanish in the limit $\lambda\to+\infty$. 
This means that for any $\varepsilon>0$, there exists $\lambda(\varepsilon)$ such that for any $\lambda>\lambda(\varepsilon)$ 
\begin{equation}
\int dx\ |\psi_n(x;\lambda)|^2V(x) < \varepsilon. 
\end{equation}
For $\lambda>\lambda(\varepsilon)$, (\ref{E'}) implies 
\begin{equation}
\frac{dE_n}{d\lambda} < \varepsilon e^{\lambda}, 
\end{equation}
which then implies 
\begin{equation}
e^{-\lambda}E_n(\lambda) < \varepsilon 
\end{equation}
for sufficiently large $\lambda$. 
Therefore, it is concluded that 
\begin{equation}
\lim_{\lambda\to+\infty} e^{-\lambda}E_n(\lambda) = 0. 
\end{equation}

For a large negative $\lambda$, as long as $V(x)$ is finite at any finite $x$, the potential wall becomes lower and lower, 
and the wave functions would spread more and more. 
The energy $E_n(\lambda)$ becomes arbitrarily small in the limit $\lambda\to-\infty$. 

When $V(x)$ is a polynomial
\begin{equation}
V(x) = \sum_{n=2}^la_nx^n, 
\end{equation}
the behavior of $E_n(\lambda)$ can be determined more explicitly. 
The result is 
\begin{equation}
E_n(\lambda) \sim \left\{ 
\begin{array}{cc}
e^{\frac{2}{l+2}\lambda}, & (\lambda\to-\infty) \\ 
e^{\frac12\lambda}. & (\lambda\to+\infty)
\end{array}
\right.
\end{equation}
This is the expected behavior from the general argument.

\end{document}